\documentstyle[12pt]{article}
  \advance\oddsidemargin  by -1in
  \advance\evensidemargin by -1in
\def\lsim{\mathrel{\rlap{\lower4pt\hbox{\hskip1pt$\sim$}}
    \raise1pt\hbox{$<$}}}         %less than or approx. symbol
\def\gsim{\mathrel{\rlap{\lower4pt\hbox{\hskip1pt$\sim$}}
    \raise1pt\hbox{$>$}}}         %greater than or approx. symbol

\def\Pom{{\bf I\!P  }}
\def\beq{\begin{equation}}
\def\endeq{\end{equation}}
\def\arr{\begin{eqnarray}}
\def\endarr{\end{eqnarray}}
\def\bm{\boldmath}

\begin{document}

\hfill{\bf \Large DFTT 28/93}

\hfill{June 1993}

\vspace{2.0cm}

\begin{center}

{\Large \bf UNITARIZATION OF STRUCTURE \\
FUNCTIONS
AT LARGE ${\bf 1/x}$
\vspace{1.0cm}\\}
{\large V.~Barone$^{a}$, M.~Genovese$^{a}$ ,
N.N.~Nikolaev$^{b,c}$,\\
E.~Predazzi$^{a}$ and B.G.~Zakharov$^{b}$ \vspace{1.0cm} \\}
{\it
$^{a}$Dipartimento di
Fisica Teorica, Universit\`a di Torino\\
and INFN, Sezione di
Torino,    10125 Torino, Italy\medskip\\
$^{b}$L. D. Landau Institute for Theoretical Physics, \\
GSP-1,
117940,
                Moscow V-334, Russia \medskip \\
$^{c}$IKP (Theorie), KFA J{\"u}lich,
5170 J{\"u}lich, Germany}
                \vspace{1.0cm}\\

{\Large \bf Abstract \bigskip\\ }

\end{center}

We discuss
the effects of the $s$-channel unitarization on the $x$ and $Q^{2}$
dependence of structure functions.
The unitarization is implemented at the level of photoabsorption
cross sections by resorting to the light--cone wave functions
of virtual photons and to the diagonalization property of the
scattering matrix in a basis of Fock states of the photon with
fixed transverse size.
Triple pomeron effects are also explicitly
taken into account. We find large unitarity
corrections to the structure functions at $x < 10^{-2}$.
The results are
in very good agreement with the existing NMC and the
preliminary HERA data.

\pagebreak

\section{ Introduction.}

The conventional QCD evolution predicts a rapid rise of the
density of partons
in the Regge limit of deep inelastic scattering,
$1/x \gg 1$ and large $Q^{2}$
(for a review
see \cite{Ba92}). The gluon distribution is predicted to increase as
\beq
xg(x,Q^{2}) \propto \exp\left\{ C\sqrt{\log
\left[{1/\alpha_{S}(Q^{2})}\right]\log\left({1/x}\right)}\right\}
\,\,\, .
\label{eq:1.1}
\endeq
and the same behaviour is expected for the sea density. This
leads to violation of the s-channel unitarity and
of the Froissart bound at large $s = Q^2/x$.

L.~V.~Gribov, Levin and Ryskin (GLR)
proposed a modified nonlinear evolution equation to account for the
recombination of partons which suppresses the increase of their densities
in the region of very small $x$ \cite{GLR83}.

In this paper we suggest to look at the unitarity
problem from a different point of view, working at the
level of cross sections. The starting point is that
the photoabsorption cross section can be derived
as an expectation value of
the interaction cross sections of the multiparton Fock states
of the virtual photon over the wave functions of such
states \cite{NZ91,NZ92}. The $s$-channel unitarity must
be imposed on the partial-wave amplitudes of interaction
of these Fock states of the photon, and this unitarization
of the partial-wave amplitudes can be reformulated in
the parton model language as the fusion of partons.
As far as the $x$ and $Q^{2}$ dependence of the total
photoabsorption cross section is concerned, the approach
\cite{NZ91,NZ92} is equivalent to the conventional
Weizs{\"a}cker-Williams reinterpretation \cite{GLDAP}
of deep inelastic scattering in terms of the (tree-level)
Compton scattering off the radiatively generated sea quarks,
but has few technical advantages: i) an exact
diagonalization of the scattering matrix in the representation
of the Fock states of the photon, ii) an easy identification
of the partial-wave amplitudes which are subject to the
s-channel unitarization.

Understanding the unitarization
effects  becomes crucial
in view of the forthcoming data from the HERA experiments,
which will go down to  $x \sim 10^{-4}$.
In this paper we shall offer a semiquantitative analysis of the
unitarity effects at large $1/x$, including the estimate of
the triple pomeron (fan diagram) contributions, the driving
term of which arises from
the $q \bar q g$ Fock component of the photon.
We extend to small $x$, including the unitarity effects,
predictions for the unitarized structure
function of the proton
$F_{2}(x,Q^{2})$, based on the dynamical generation
of the gluon and sea distributions starting with the
three constituent quarks \cite{BGNPZ91,BGNPZ93a,BGNPZ93b}.

%                    Section   2

\section{The color dipole cross section as a subject of
unitarization}

At $1/x \gg 1$ photoabsorption
can be treated as the scattering on the nucleon of the $q\bar{q}$
pairs the virtual photon transforms into at large
distances $\Delta z \sim 1/m_{_N}x \gg R_{N}$
upstream the target nucleon (Fig.~1). In this limit the transverse size
$\mbox{\bm $\rho$}$ of the $q\bar{q}$ pair
and the Sudakov variable
$z$ , i.e. the fraction of the photon's light-cone
momentum carried by one of the quarks of the pair ($0 <z<1$),
are frozen in the scattering process. The scattering matrix
becomes diagonal in the $(\mbox{\bm $\rho$},z)$ representation,
which allows one to write
the  transverse (T) and
 longitudinal (L) virtual photoabsorption cross sections
as the quantum mechanical expectation value
\beq
\sigma_{T,L}(Q^{2})=
\int_{0}^{1} dz\int d^{2}\mbox{\bm$\rho$}\,\,
\vert\Psi_{T,L}(z,\rho)\vert^{2}\sigma_0(\rho)\,\,\,.
\label{eq:2.1}
\endeq
Here $\sigma_{0}(\rho)$ is the interaction cross
section for the $q\bar{q}$ color dipole of size
{\mbox {\bm $\rho$}}, given by
\cite{NZ91}
\beq
\sigma_0(\rho)={16 \over 3 }\alpha_{S}(\rho)\int
{ d^{2}{\bf k}\,V({\bf k})[1-\exp(i{\bf k}\cdot \mbox{\bm$\rho$})]
\over
({\bf k}^{2} + \mu_{G}^{2})^{2} }
\alpha_{S}({\bf k}^{2}) \,\,\,.
\label{eq:2.2}
\endeq
In (\ref{eq:2.2}) ${\bf k}$ is the transverse momentum of the exchanged
gluons (Fig.1a),  $\,\mu_{G} \approx 1/R_{c}$
is the effective mass of gluons introduced
so that color forces do not propagate beyond the confinement
radius $R_{c}$. The gluon--gluon--nucleon vertex
function $V({\bf k}) =
1-{\cal F}_{ch}(3{\bf k}^{2})$,
where ${\cal F}_{ch}(q^{2})$ is the charge form factor of the proton.

The wave functions of the $q\bar{q}$ fluctuations of the photon were
derived in \cite{NZ91} and read
\beq
\vert\Psi_{T}(z,\rho)\vert^{2}={6\alpha_{em} \over (2\pi)^{2}}
\sum_{1}^{N_{f}}e_{f}^{2}
\{[z^{2}+(1-z)^{2}]\varepsilon^{2}K_{1}(\varepsilon\rho)^{2}+
m_{f}^{2}K_{0}(\varepsilon\rho)^{2}\}\,\,,
\label{eq:2.3}
\endeq
\beq
\vert\Psi_{L}(z,\rho)\vert^{2}={6\alpha_{em} \over (2\pi)^{2}}
\sum_{1}^{N_{f}}e_{f}^{2}\,\,
(4Q^{2})\,z^{2}(1-z)^{2}K_{0}(\varepsilon\rho)^{2}\,\,,
\label{eq:2.4}
\endeq
where $K_{\nu}(x)$
are modified Bessel functions,
$\varepsilon^{2}=z(1-z)Q^{2}+m_{f}^{2}$ and
$m_{f}$ is the quark mass.
Since either $V({\bf k})$ and/or the factor
$[1-\exp(i{\bf k}\cdot\mbox{\bm $\rho$})]$
vanish as $\vert {\bf k}\vert \rightarrow 0$, the dipole
cross section
$\sigma_0(\rho)$ is free from infrared
divergences.
At small $\rho$
\beq
\sigma_{0}(\rho) \propto \rho^{2}\,\alpha_{S}(\rho)
\log\left[1/\alpha_{S}(\rho)\right]\,\,,
\label{eq:2.5}
\endeq
and vanishes as $\rho \rightarrow 0$.

Eq.~(\ref{eq:2.2}) gives the driving, constant term of the QCD pomeron
contribution to the dipole cross section. The $x$ dependence of
the dipole cross section $\sigma_{0}(x,\rho)$ comes from the
$t$-channel iteration of the gluon-exchange between the two
gluons in the diagram of Fig.1b. These higher order contributions
can be understood as the QCD evolution of the density of gluons
in the nucleon, or interactions of the higher $q\bar{q}g_{1}...g_{n}$
Fock states of the photon \cite{NZ93}. The convenient way to
incorporate this
$x$-dependence in is as follows:

The driving term $G(x,Q^{2})$ of the distribution of
soft, $x \ll 1$, gluons
with $0< {\bf k}^{2} < Q^{2}$, generated perturbatively
from the nucleon, is given by \cite{BGNPZ93a,BGNPZ93b}
\beq
G(x,Q^{2})=
\frac{4}{\pi} \frac{1}{x}
\int_{0}^{Q^{2}} \frac{ {\rm d}{\bf k}^2 \, {\bf k}^2}{
({\bf k}^{2}+\mu_{G}^{2})^{2} }
\, \alpha_{S}({\bf k}^{2}) \,
V({\bf k})\,.
\label{eq:2.11}
\endeq
Comparing eq.~(\ref{eq:2.2}) and eq.~(\ref{eq:2.11})
we can rewrite the dipole cross section as
\beq
\sigma_{0}(\rho)= {4 \pi^2 \over 3}\,\alpha_{S}(\rho)\, \int
\! \frac{ {\rm d}{\bf k}^2}{{\bf k}^{4}}\,
(1-{\rm e}^{i{\bf k}\cdot\mbox{\bm $\rho$}}) \,
\frac{{\rm d}(x  G(x,k^{2}))}{{\rm d}
\log {\bf k}^2}
\,\,.
\label{eq:2.13}
\endeq
The generalized QCD ladder diagrams
of Fig.1b are summed by replacing the
driving term of the gluon flux $G(x, Q^2)$ with the
full GLDAP--evolved gluon distribution $g(GLDAP,x,Q^{2})$
of our previous work \cite{BGNPZ93a,BGNPZ93b},
{\it i.e.}
\beq
\sigma_{0}(x,\rho)= {4 \pi^2\over 3} \,\alpha_{S}(\rho)\, \int
\! \frac{ {\rm d}{\bf k}^2}{{\bf k}^{4}}\,
(1-{\rm e}^{i{\bf k}\cdot \mbox{\bm $\rho$}}) \,
\frac{{\rm d} \, x g(GLDAP,x,k^{2})}{{\rm d}\log {\bf k}^{2}}
\,\,.
\label{eq:2.15}
\endeq
This GLDAP dipole cross section (\ref{eq:2.15}) must then be used in
calculation of the photoabsorption cross section (\ref{eq:2.1})
and, eventually, of the GLDAP structure function using the definition
$F_{2}(x,Q^{2})= Q^{2}\sigma_{\gamma^{*}N}(x,Q^{2})/
4\pi^{2}\alpha_{em}$.
Since
$\sigma(x,\rho) \sim \rho^{2}\alpha_{S}(\rho)g(x,Q^{2})$,
in view of Eq.(\ref{eq:1.1}) at small $x$ it violates both
the Froissart bound and the $s$-channel unitarity.

Now, let us
identify the corresponding partial waves using the conventional
impact parameter representation for the scattering amplitude
\beq
f_{0}({\bf q})=2i\, \int d^{2}{\bf b}\,
\Gamma_{0}({\bf b})\exp(-i{\bf b}\cdot {\bf q})
\,\, .
\label{eq:3.1}
\endeq
For the predominantly
imaginary scattering amplitude
$f_{0}({\bf q})=i\sigma_{0}(x,\rho)\exp(-B_{0}{\bf q}^{2}/2)$ the
profile function (partial-wave amplitude) $\Gamma_{0}({\bf b})$
of the GLDAP dipole cross section (9) equals
\beq
\Gamma_{0}({\bf b})=\frac{\sigma_{0}(x,\rho)}{ 4\pi B_0}
\exp\left(-\frac{{\bf b}^{2}}{2 B_0}\right)
\label{eq:3.3}
\endeq
and
violates the $s$-channel
unitarity constraint $\Gamma({\bf b}) \leq 1$
when $\sigma_{0}(x,\rho)$ becomes large.
 (By geometrical considerations, the diffraction slope $B_0$
can be related to the conventional hadronic slope for $\pi p$
scattering $B_{\pi p}$
through $B_0 = B_{\pi p}/2 + \rho^2/4$).

We encounter, in fact, the as yet
unsolved problem of unitarization of the rising
strong interaction cross section.
Two different schemes are commonly adopted for construction
of the unitarized partial-wave
amplitude $\Gamma({\bf b})$ : {\it i)} the
eikonal unitarization \cite{A67}
\beq
\Gamma({\bf b})=1-\exp\left[-\Gamma_{0}({\bf b})\right]\,,
\label{eq:3.4}
\endeq
and the ${\cal K}$ matrix  unitarization \cite{A64}
\beq
\Gamma({\bf b})={\Gamma_{0}({\bf b}) \over 1+\Gamma_{0}({\bf b}) }
\,.
\label{eq:3.7}
\endeq
In both cases as the bare profile function $\Gamma_{0}({\bf b})$
rises, the
unitarized one $\Gamma({\bf b})$ tends to the black disk limit
$\Gamma({\bf b}) \rightarrow 1$.

The unitarized total cross sections
$\sigma(x,\rho)=2\int d^{2}{\bf b}\,\Gamma({\bf b})$
reads
\beq
{\rm eikonal:}\;\; \sigma(x,\rho) = 4 \pi B \left [
\log (\eta(x, \rho)) + E_1(\eta) + \gamma \right ]\,;
\label{eq:3.5}
\endeq
\beq
{\rm {\cal K} \;matrix:}\;\;
\sigma(x,\rho)=4\pi B \log(1+\eta(x,\rho))\,,
\label{eq:3.8}
\endeq
where $\gamma$ is the Euler--Mascheroni constant, $E_1$ is the
integral exponential function.  The quantity
$\eta(x,\rho) = \sigma_0(x,\rho)/4 \pi B_0$
shown in Fig.2, controls the
effect of the unitarization: $ \sigma(x,\rho) \simeq
\sigma_0(x,\rho)$ at $\eta(x, \rho) \ll 1$, and
at $\eta(x,\rho) \gg 1$ the unitarization suppresses the
cross section, $\sigma(x,\rho) \ll \sigma_0(x,\rho)$.
This plot shows that the unitarization effects are important
only at large $\rho$.
The difference between the two cross sections (\ref{eq:3.5}) and
(\ref{eq:3.8}) can be taken as a measure of the theoretical
uncertainty
on the unitarization procedure. For convenience
we shall adopt hereafter the ${\cal K}$--matrix procedure,
the eikonal unitarization gives similar, if somewhat weaker
unitarizations effects.

%----------------------------------

%                   Section 4

\section{Unitarity and the triple pomeron contribution}

To the leading order in the $s$--channel unitarization
$\Gamma({\bf b})=\Gamma_{0}({\bf b})-\chi \Gamma_{0}({\bf b})^{2}/2$,
where $\chi=1(2)$ for the eikonal ({\cal K}-matrix)
unitarization, and the unitarized photoabsorption
cross section reads
\beq
\sigma_{\gamma^* N}(x, Q^2) = 2\int d^{2}{\bf b}\,\Gamma({\bf b})=
\langle \sigma_0 \rangle - \chi
\langle \int d^{2}{\bf b} \Gamma_{0}({\bf b})^{2}\rangle\,,
\label{3pom.1}
\endeq
where we have denoted by
$\langle \cdot \rangle$ the average over the
photon wave function. The  first term in eq.~(\ref{3pom.1})
gives the GLDAP cross section and the
conventional GLDAP structure
function, the second term is the unitarity (shadowing) correction,
which in principle is a nonlinear functional of the GLDAP
cross section.
The unitarity correction
can be related to the cross section of the forward
diffraction dissociation of the virtual photons
$\gamma^{*}+p \rightarrow X+p$, where
$X=q\bar{q}$, as follows \cite{NZ92} :
\beq
\sigma_{D}(\gamma^{*}\rightarrow q\bar{q})=
\langle \int d^{2}{\bf b} \Gamma_{0}({\bf b})^{2}\rangle\
=\int {\rm d}t\,{\rm d} M^2
\frac{{\rm d} \sigma_D(\gamma^{*}\rightarrow q\bar{q})}
{{\rm d} M^2 {\rm d}t}
=\int {\rm d} M^2
\left.\frac{{\rm d} \sigma_D(\gamma^{*}\rightarrow q\bar{q})}
{B_{D}{\rm d} M^2 {\rm d}t}
\right|_{t = 0}\,\, ,
\label{eq:4.2}
\endeq
\beq
\left.\frac{{\rm d} \sigma_D(\gamma^{*}\rightarrow q\bar{q})}
{{\rm d}t}
\right|_{t = 0}=\int {\rm d} M^2
\left.\frac{{\rm d} \sigma_D(\gamma^{*}\rightarrow q\bar{q})}
{{\rm d} M^2 {\rm d}t}
\right|_{t = 0}= \frac{1}{16 \pi} \langle \sigma_{0}^{2} \rangle \, .
\label{3pom.2}
\endeq
Here $B_{D}$ stands for the diffraction slope of the diffraction
dissociation.
The driving term of diffraction dissociation is excitation
of the $q \bar q$ component of the photon, which
has the $\propto 1/(Q^{2}+M^{2})^{2}$ mass spectrum and gives
the $x$-independent shadowing term. Diffraction excitation of
the higher $q\bar{q}g_{1}...g_{n}$ Fock states of the photon
gives rise to the triple pomeron component of the mass
spectrum at $M^{2} \gg Q^{2}$:
\beq
\left. \frac{{\rm d} \sigma_D(\gamma^{*}\rightarrow q\bar{q}g...)
}{{\rm d}t {\rm d} M^2}\right|_{t=0}
= G_{3 \Pom} \, \frac{1}{M^2}\,,
\label{3pom.3}
\endeq
and to the logarithmically, $\propto \log(s) \propto log(1/x)$ ,
rising shadowing component.

We shall present a brief discussion of the effects of the
$q\bar{q}g$ Fock state, which is the driving term of the triple-pomeron
mass spectrum. We apply to the $q \bar q g$ system the same
wave function formalism already used for the $q \bar q$ component:
\beq
\int {\rm d} M^2\,
\left. {
{\rm d} \sigma_{D}(\gamma^{*}\rightarrow q\bar{q}g)
\over
{\rm d}t {\rm d} M^2} \right|_{t=0}
 =
\frac{1}{16 \pi}
\int_{x}^{1} {\rm d} z_g \int {\rm d} z \int {\rm d}^2
{\bf r} \int {\rm d}^2 \mbox{\bm $\rho$} \,
\vert \Phi(\mbox{\bm $r$, $R$, $\rho$}, z, z_g) \vert^2
\sigma_0(r,R,\rho)^2 \,.
\label{3pom.6}
\endeq
Here we have introduced the three-particle wave function $\Phi$
and the three-particle interaction cross section
$\sigma_{0}(r,R,\rho)$, where
${\bf r}$, ${\bf R}$ and
$\mbox{\bm $\rho$}$ are the $g \bar q$, $q g$ and $q \bar q$
separations, respectively, and $z_g$ is the fraction of the photon's
light--cone momentum carried by the gluon.
It is possible to show \cite{NZ93}
that the leading contribution to the triple
pomeron component comes from the ordering
$1/Q^2 \ll \rho^2 \ll r^2 \sim R^2$. In this case the
$q\bar{q}$-pair of the $q\bar{q}g$ Fock state can be
treated as the pointlike color-octet charge, so
that the three-particle cross section $\sigma_0(r, R, \rho)$
reduces to
\beq
\sigma_0(r,R,\rho) \rightarrow \frac{9}{4} \sigma_0(r)\,,
\label{eq:3pom.5}
\endeq
where $9/4$ is the familar ratio of the octet and the triplet
strong couplings. The three-particle wave function takes on
the factorized form
\beq
\vert \Phi \vert^2 = \frac{2}{3 \pi^2} \frac{1}{z_g} \alpha_s(\rho)
\vert \Psi(z,\rho) \vert^2  \frac{\rho^2}{r^4} \, {\cal F}(\mu_G r)\,,
\label{3pom.7}
\endeq
which holds when the gluon of the
$q\bar{q}g$ Fock state {\sl participates} in the interaction,
so that only the diagram of  Fig.~1c (and not the
one of Fig.~1d) should be taken into account and the
corresponding three-particle cross section is reduced to
half of (\ref{eq:3pom.5}), we refer to \cite{NZ93} for a
derivation and more details.
Here $\Psi(z,\rho)$ is the wave function of the $q \bar q$ state
and ${\cal F}(x) = (x^2 K'_1(x))^2$ satisfies ${\cal F}(0) = 1$ and
${\cal F}(x) \sim \exp(- 2 x)$ at large $x$.

Since at $M^{2} \gg Q^{2}$ one has $dM^{2}/M^{2} = dz_{g}/z_{g}$,
the $1/z_{g}$ dependence of the wave function (\ref{3pom.7}), which
has its origin in the spin 1 of gluons, leads to
the mass spectrum of the form (19) and to the $\propto \log(1/x)$
dependence of the shadowing term. Using the wave function (22) and
the cross section (21) it can be shown that the
triple-pomeron component of the diffraction
dissociation cross section (20) satisfies an approximate
factorization property:
\beq
\left. \frac{{\rm d} \sigma_{D}(\gamma^{*}\rightarrow X)}
{{\rm d}t {\rm d} M^2} \right|_{t=0}
\approx \sigma_{\gamma^{*}N} A_{3\Pom} {1 \over M^{2}}
\label{eq:---}
\endeq
where $A_{3\Pom}$ only weakly depends on $x$, $Q^{2}$
and on the flavour of quarks in the $q\bar{q}$ state.
This approximate factorization allows one to write down the
unitarized structure function in the form
\beq
F_{2}(x,Q^{2})=F_{2}(GLDAP,x,Q^{2})
\left(1-\Delta Sh(q\bar{q})-
\chi{A_{3\Pom}\over B_{3\Pom}}\log(1+{x_{0} \over x})\right)\, .
\label{eq:"""}
\endeq
Here we singled out the shadowing
correction from excitation of the $q\bar{q}$ Fock state, which
has a strong flavour dependence,
$\Delta Sh(q\bar{q}) \propto 1/m_{f}^{2}$, and vanishes
$\propto 1/Q^{2}$ for the longitudinaly polarized photons
\cite{NZ91,NZ92,BGNPZ93c}.
In Eq.~(24) $B_{3\Pom}$ stands for the diffraction slope of
the diffraction dissociation in the triple-pomeron region,
and $x_{0} \approx 0.1$ corresponds to the usual definition
of the diffraction excitation as interactions in which the target
nucleon receives little recoil and emerges in the final state
separated from the virtual photon's debris by the rapidity
gap $\Delta \eta =\log(1/x_{0}) \gsim 2$. Diffraction
excitation of the $q\bar{q}$ Fock state dominates the mass
spectrum at $M^{2} \sim Q^{2}$ and is the counterpart of
excitation of resonances in the hadronic scattering,
which has the diffraction slope $B_{D}$ close to that
of the elastic scattering, so that
here we take $B_{D} = 10\,(GeV/c)^{-2} \approx B_{\pi N} $, whereas
in the triple-pomeron region $B_{D}=B_{3\Pom}\approx B_{\pi N}/2$
(for the review see \cite{Chapin}).
More complete treatment of the shadowing, in which one
describes the diffraction dissociation in terms of the structure
function of the pomeron, does not change much the above
simple estimate of the triple-pomeron effect.

\section{Discussion of the results }

In our approach to DIS at small-$x$ the fundamental quantity is
the dipole cross section $\sigma_{0}(x,\rho)$. We assume that
at large $\rho \sim R_{N}$ it is practically flat as a function of $x$,
which gives a smooth connection with the slow
rise of the hadronic cross sections \cite{DL}. This
corresponds also to our treatment of the glue in nucleons
at $Q^{2} \leq Q_{0}^{2}$ as a radiative effect,
starting from three valence quarks at the
spectroscopic scale, as explained in detail in our previous papers
\cite{BGNPZ93a,BGNPZ93b}. Beyond
$Q_{0}^{2}$ we calculate $g(x,Q^{2})$ using the conventional
$GLDAP$ evolution. At small $Q^{2}$  we use the freezing
coupling constant $\alpha_s(Q^2 \le Q_f^2) = \alpha_s(Q_f^2)$,
where $Q_f^2$ is a scale of the same order in magnitude
as $Q_0^2$. Also, at $Q^{2}\leq Q_{0}^{2}$  we
neglect the splitting of gluons and the momentum flow
from glue to sea.
The absolute normalization of the sea structure function
is determined by the absolute normalization of the
dipole cross section $\sigma_{0}(\rho)$, which is fixed by the
confinement radius $R_{c}=1/\mu_{G}$ and the frozen strong
coupling $\alpha_{S}(Q_{f}^{2})$. These two parameters
$\mu_{G}$ and $Q_{f}$ are
strongly correlated when we ask for simultaneous description of
the $\pi N$ total cross section.
This procedure corresponds to the minimal infrared
regularization and has been
already applied with success to various structure function calculations
\cite{BGNPZ91,BGNPZ93a,BGNPZ93b,BGNPZ93c}.
The choice of parameters used in the present calculations
is: $Q_0^2 = 1.0 \, GeV^2/c^2$, $Q_f^2
= 0.5 \, GeV^2/c^2$, $\mu_G = 300 \, MeV/c^2$.

With the above set of parameters, in the region of small $Q^2$ we
find $A_{3 \Pom} \simeq 0.15 \, GeV^{-2}$, to be
compared with the experimental result for the diffraction dissociation
of real photons $A_{3 \Pom} = 0.16 \, GeV^{-2}$ \cite{Chapin}.
As stated above, our calculations show that
$A_{3\Pom}$ changes little, $\lsim 10\%$, with
$Q^{2}$ and on the flavour of quarks.

We {\sl predict} the GLDAP and unitarized structure functions
with {\sl absolute} normalization.
The predictions for the unitarized structure function
$F_2(x,Q^2)$ at different
$x$ and $Q^2$ are shown in Fig.~3 and Fig.~4.
compared to the existing NMC data \cite{NMC}
at $x= 0.025$ and to the preliminary
HERA data \cite{H1,ZEUS} at $x=1.5 \cdot 10^{-3}, 4 \cdot 10^{-4}$.
The unitarity correction is numerically large and makes
the rise of the structure function with $1/x$ much less
steep. Considering the accuracy of the NMC data, the
shadowing correction is rather large already at
$x \sim 10^{-2}$, where the effect of shadowing of the $q\bar{q}$
Fock state of the photon and the triple-pomeron contribution
are of comparable magnitude. At smaller $x$ the triple-pomeron
component starts taking over, but the $q \bar {q}$ component also
rises slowly as it is proportional to
$ \langle \sigma(x,\rho)^{2} \rangle$.
We find very good agreement
between our predictions for the unitarized structure function
and the experiment. Notice, that the triple pomeron
component tends to dominate
as $x$ decreases.

In Fig.~5 we compare our prediction for $F_2$ in the
very small $x$
region with some parametrizations \cite{MRS,DG}, based on
fits to the previously published experimental data on
structure functions. At very small $x\sim 10^{-4}$ our
structure
functions follow an approximate law $F_{2}(x,Q^{2})
\propto (1/x)^{\Delta(Q^{2})}$, where the exponent
$\Delta(Q^{2})\approx 0.21$ at $Q^2 = 4 \, GeV^2/c^2$
and
$\Delta(Q^{2})\approx 0.311$ at $Q^2 = 15 \, GeV^2/c^2$.

In Fig.~6 we present the effects
of the unitarization on the charm component of
the structure function. We take $m_{c}=1.7\,GeV/c^{2}$.
Here the shadowing of the $q\bar{q}$
Fock state is negligible, and the unitarity correction
is completely
dominated by the triple-pomeron component.
The situation with the ratio $R=\sigma_{L}/\sigma_{T}$ is similar:
the $\Delta Sh(q\bar{q})$ component is small, and the
unitarization of the longitudinal structure function is
dominated by the triple pomeron component. We find that
$R$ is quite structureless and depends very weakly on $Q^{2}$
at very small $x$. At $x \lsim 10^{-2}$ we predict
$R\approx 0.20-0.24$.

After this work was completed, there appeared an estimate
of the unitarity effects by Askew et al. \cite{Askew}, who use
a very different formalism. In \cite{Askew} the parameter
which controls the shadowing is the radius $R_{g}$ of the
region of the proton in which the gluons are concentrated.
With $R_{g}=2\,GeV^{-1}$, which is significantly smaller
than the proton's size, Askew et al. find unitarity
corrections close to ours at $x \lsim 10^{-3}$, and
have no shadowing at $x \sim 10^{-2}$.

\section{Conclusions}

We have presented a simple approach to the unitarization of
rising structure functions at small $x$. The shadowing term
in the unitarized structure function is dominated by the
triple-pomeron term, which is approximately flavour- and
$Q^{2}$--independent. We find large shadowing corrections
already at relatively
large $x\sim 10^{-2}$. We find that the dynamical generation
of the glue and sea from the valence quarks, which predicts
the small-$x$ structure
functions with {\sl absolute} normalization, is in good
agreement
with the preliminary data from HERA \cite{H1,ZEUS}.

\pagebreak

\pagebreak
{\bf \Large Figure captions:}
\begin{itemize}

\item[Fig.1]

- a) Driving term of the QCD pomeron, ~b) Generalized ladder
diagrams for the QCD pomeron, ~c) Interaction of the $q\bar{q}g$
Fock state of the photon which gives the rising contribution
to the total cross section, ~d) Interaction with the color-octet
$q\bar{q}$ component of the $q\bar{q}g$ state which does not
contribute to the cross section growth.

\item[Fig.2]

- The unitarization parameter $\eta(x,\rho)$.

\item[Fig.3]

- Predictions for the GLDAP (dashed curve) and unitarized
(solid curve) structure functions at small $x$ compared to
the NMC \cite{NMC} data at $x=0.025$ and the preliminary
H1 \cite{H1} and ZEUS \cite{ZEUS}
data at small $x$. The dot-dashed curve shows the effect
of shadowing of the $q\bar{q}$ Fock state of the photon.

\item[Fig.4]

- Our prediction for the $x$ dependence of the unitarized
structure function at $Q^{2}=15\,(GeV/c)^{2}$ compared to
the NMC \cite{NMC}, H1 \cite{H1} and ZEUS \cite{ZEUS}
data.

\item[Fig.5]

- Comparison of our prediction for the small-$x$ behaviour
of $F_{2}(x,Q^{2})$ (solid curve) with some of the recent
MRS \cite{MRS} and DGJLP \cite{DG}
parametrizations based on fits to the published data.

\item[Fig.6]

- Our prediction for the GLDAP (dashed curve) and the
unitarized (solid curve) charm distribution at
$Q^{2}=15\,(GeV/c)^{2}$. The effect of shadowing of
the $c\bar{c}$ Fock state of the photon is shown by
the dot-dashed curve.

\end{itemize}
\end{document}